\documentclass[twocolumn,showpacs,amsmath,amssymb]{revtex4}

\usepackage{graphicx}
\usepackage{dcolumn}
\usepackage{bm}

\begin{document}

\title{Application of the Hilbert-Huang Transform to the Search for Gravitational Waves}
\author{Jordan B. Camp}
\email{camp@lheapop.gsfc.nasa.gov}
\author{John K. Cannizzo}
\altaffiliation[Also at ]{Physics Department, University of Maryland, Baltimore County, Baltimore, Maryland 21250}
\author{Kenji Numata}
\altaffiliation[Also at ]{Department of Astronomy, University of Maryland, College Park, Maryland 20742}
\affiliation{Laboratory for Gravitational Physics, Goddard Space Flight Center, Greenbelt, Maryland 20771}
\date{\today}
\begin{abstract}
We present the application of a novel method of time-series analysis, the Hilbert-Huang Transform, to the search for gravitational waves. This algorithm is adaptive and does not impose a basis set on the data, and thus the time-frequency decomposition it provides is not limited by time-frequency uncertainty spreading. Because of its high time-frequency resolution it has important applications to both signal detection and instrumental characterization. Applications to the data analysis of the ground and space based gravitational wave detectors, LIGO and LISA, are described.
\end{abstract}
\maketitle

{\it Introduction ---} A vital area of gravitational wave (GW) research is the development of data analysis algorithms that are able to detect non-linear, transient signals.  The data analysis effort is especially relevant as the Laser Interferometer Gravitational Wave Observatory (LIGO) \cite{Barish1999,Abramovici1992} undertakes a one-year science run at full design sensitivity.  Also, the Laser Interferometer Space Antenna (LISA) \cite{LISA} is engaged in the development of analysis algorithms to search for its candidate sources.

The Hilbert Huang Transform (HHT) \cite{Huang1998} presents a fundamentally new approach to the analysis of time series data. Its essential feature is the use of an adaptive time-frequency decomposition that does not impose a fixed basis set on the data, and therefore it is not limited by the time-frequency uncertainty relation characteristic of Fourier or Wavelet analysis. This leads to a highly efficient tool for the investigation of transient and non-linear features. Applications of the HHT include materials damage detection \cite{Yang2004} and biomedical monitoring \cite{Novak2004,HHT}. Because General Relativity is an inherently non-linear theory, and because LIGO, LISA, and other GW detectors produce a great variety of non-linear and transient signals, the HHT has the promise of being a powerful new tool in the search for gravitational waves. This article describes the application of the HHT to GW data analysis.

The HHT proceeds in two steps. First, the process of Empirical Mode Decomposition (EMD) reduces the time-series under analysis into components, known as Intrinsic Mode Functions (IMFs), thereby "sifting" or separating out the different frequency scales of the data. The IMFs when summed reproduce the original time series. The sifting is done adaptively, with no a priori structure imposed on the data.

The IMFs have a vertically symmetric and narrowband form that allow the second step of the HHT to be applied: the Hilbert transform of each IMF. As explained below, the Hilbert transform obtains the best fit of a sinusoid to each IMF at every point in time, identifying an instantaneous frequency (IF), along with its associated instantaneous amplitude (IA). The IF and IA provide a time-frequency decomposition of the data that is highly effective at resolving non-linear and transient features.

{\it Instantaneous Frequency ---}
The IF is generally obtained from the phase of a complex signal $z(t)$ which is constructed by analytical continuation of the real signal $x(t)$ onto the complex plane \cite{HHT}. By definition, the analytic signal is 
\begin{equation}
z(t) = x(t) + i y(t)
\end{equation}
where y(t) is given by the Hilbert Transform:
\begin{equation}
y(t)=\frac{1}{\pi}P\!\!\int_{-\infty}^{\infty}\frac{x(t')}{t-t'}dt'
\end{equation}
(Here $P$ denotes the principal Cauchy value.) The amplitude and phase of the analytic signal are defined in the usual manner: $a(t)=|z(t)|$ and $\theta(t)=\arg{z(t)}$ .The analytic signal represents the time-series as a slowly varying amplitude envelope modulating a faster varying phase function \cite{TF}. The IF is then given by $\omega(t)=d\theta(t)/dt$, while the IA is $a(t)$. We emphasize that the IF, a function of time, has a very different meaning from the Fourier frequency, which is constant across the data record being transformed. Indeed, as the IF is a continuous function, it may express a modulation of a base frequency over a small fraction of the base wave-cycle. Conditions on the form of $x(t)$ that allow the application of this procedure are: 1) vertical symmetry with respect to the local zero mean, and 2) the same number of zero crossings and extrema \cite{Huang1998}.

A time-frequency analysis of a frequency modulated signal serves to illustrate the concept of instantaneous frequency. An important GW source for LISA is the inspiral of supermassive black holes (SMBH). The GW waveform from this source has a base frequency which chirps as the masses approach, and is also phase and amplitude modulated due to spin-orbit coupling \cite{Vecchio2004}. The time evolution of the base frequency of the GW from the inspiral is given to first order by \cite{Blanchet}:
\begin{equation}
f=\frac{1}{8\pi}\left( \frac{GM_{\rm t}}{c^3}\right)^{-\frac{5}{8}}\left( \frac{\mu}{M_{\rm t}}(t_{\rm 0}-t)\right)^{-\frac{3}{8}}
\label{eqn:merger}
\end{equation}
where $\mu=m_1m_2/(m_1+m_2)$ is the system reduced mass, $M_{\rm t}$ is the system total mass, $G$ is Newton's constant, $c$ is the speed of light, and $t_0$ is the coalescence time.

\begin{figure}
\includegraphics*[width=\linewidth]{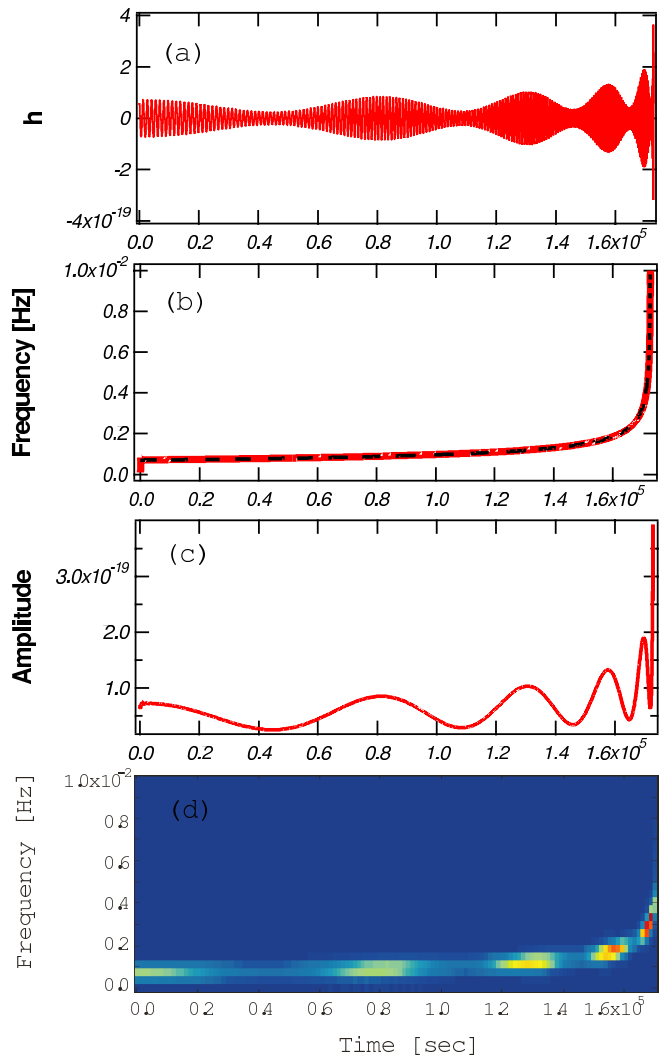}
\caption{\label{fig:SMBH}Comparison of IF and FFT-based spectrogram of GW signal from SMBH inspiral with spin-orbit coupling. (a) GW waveform. (b),(c) IF and IA show sharp detail of frequency and amplitude modulation. The dotted line on (b) plots Eq.(\ref{eqn:merger}). (d) FFT spectrogram is blurred because of time-frequency spreading and spurious harmonics.}
\end{figure}

Fig.\ref{fig:SMBH}-(a) shows the waveform from an SMBH binary inspiral with masses $m_1$ and $m_2$ of $10^6$ and $10^5$ solar mass, and spins $s_1$ and $s_2$ of $0.95 m_1^2$ and $0.95 m_2^2$. This waveform is used as input to the HHT and FFT-based spectrogram to perform the time-frequency analysis, shown in Figs.\ref{fig:SMBH}-(b), -(c), -(d). It is clear that the extraction of the IF and IA provides high time-frequency resolution of the waveform, showing detail of the chirp as predicted by Eq.(\ref{eqn:merger}) (dotted line in Fig.\ref{fig:SMBH}-(b)), and also amplitude modulation due to the spin-orbit coupling. In contrast, the decomposition of the chirp through the FFT is seen to be limited by the time-frequency uncertainty relation: $\Delta t \Delta f \sim1$, where $\Delta t$ and $\Delta f$ are the time interval and (Fourier) frequency bandwidth. The decoupling of the modulations of the IF and IA provided by the HHT allows a detailed look at the amplitude and frequency evolution of the waveform, and will aid significantly in determining the SMBH binary parameters.

{\it Empirical Mode Decomposition ---}
The waveform of the previous example has an approximate symmetry about zero that satisfies the conditions listed above and allows meaningful application of the Hilbert Transform for extraction of the IF. To express an arbitrary time-series in this form the process of EMD is applied \cite{Huang1998}.

\begin{figure}
\includegraphics*[width=\linewidth]{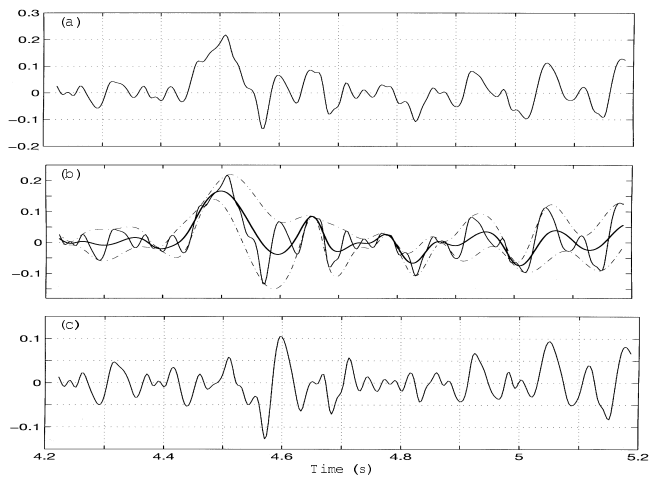}
\caption{\label{fig:EMD}Illustration of Empirical Mode Decomposition (from Ref.\cite{Huang1998}). (a) time series. (b) average of envelope formed by fitting extrema. (c) subtraction of average from time-series. Because the result is not vertically symmetric, the waveform will be re-sifted until an IMF is identified.}
\end{figure}

The EMD process consists of forming an envelope about the extrema of the data by cubic spline fitting, and then subtracting the average of the envelope from the data. The extrema of remainder are then fitted, and the process is repeated as many times as necessary to obtain a waveform that is symmetric about zero mean (within a predetermined tolerance). Once this has occurred, the waveform is labeled IMF1 and is subtracted from the original time series, removing the highest frequency content, and allowing the remainder to be sifted again to obtain the next IMF. This procedure is illustrated in Fig.\ref{fig:EMD}.

The sifting identifies and removes components of the data first at the highest frequencies, then down in frequency to the lowest trends. For white noise the EMD acts as a dyadic filter, so that the central frequency associated with IMF$n$ is of order $f_{\rm samp}/2^n$ where $f_{\rm samp}$ is the data sampling frequency \cite{Huang2004}. A consequence of the sifting is that the IMFs are vertically symmetric about zero mean (although both the frequency and amplitude of an IMF vary in time.) The narrowband and symmetric form of the IMFs are essential to allow the Hilbert Transform to be applied in a meaningful way \cite{Huang1998}. 

\begin{figure}
\includegraphics*[width=\linewidth]{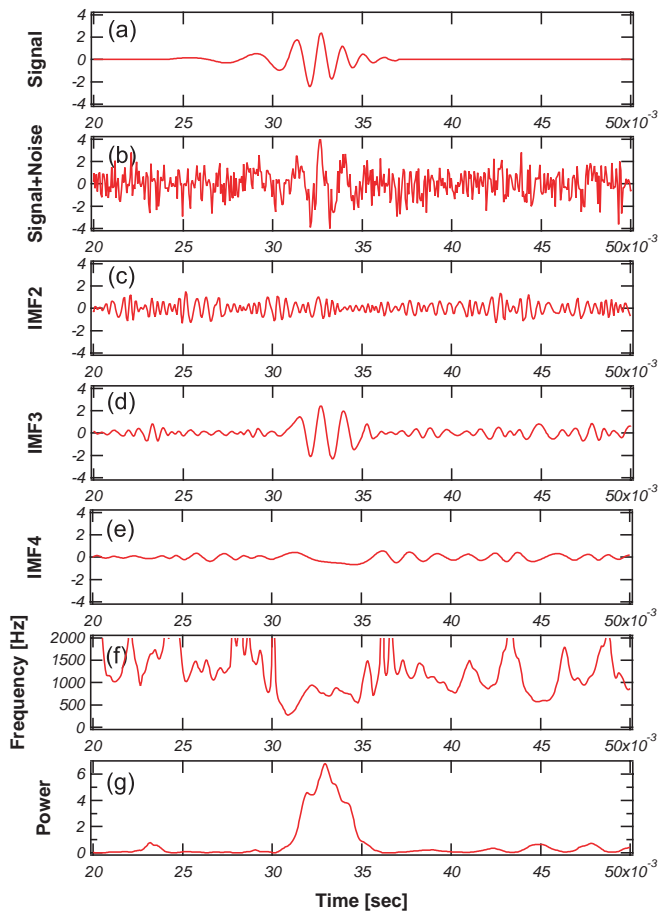}
\caption{\label{fig:merger}HHT of 20 solar mass black hole binary merger signal in white noise. (a) merger signal. (b) signal in noise with SN=10. (c),(d),(e) IMFs 2,3,4. (f),(g) instantaneous frequency and power derived from the Hilbert Transform of IMF3 show the time-frequency-power structure of the merger.}
\end{figure}

{\it EMD and the Hilbert Transform: The HHT---}
To illustrate the application of the full HHT to GW analysis, we look at the identification of a signal in white noise (Fig.\ref{fig:merger}). As an example relevant to LIGO analysis, where strong signals are not expected, we consider here a weak signal relative to the noise.  An issue for the HHT is that low signal to noise tends to degrade the IF and IA identification, as the signal extrema becomes distorted. A rough measure of this tendency was determined from simulations by injecting sine waves of varied signal to noise, frequency and duration into white noise. This showed $\Delta f/f \sim 2\sqrt{\delta t/5 {\rm msec}}/{\rm SN}$, where $\Delta f$ is the extracted frequency width of a periodic wave of frequency $f$ and duration $\delta t$ in white noise, with signal to noise SN \cite{SNdef}. Thus in searching for signals with ${\rm SN}<20$ the HHT works best for $\delta t<20 {\rm msec}$.

As an example we inject a 20 solar mass black hole binary merger and ringdown signal \cite{Baker2006} with SN=10 and a time duration of 5~msec into white noise at 16~kHz sampling rate. The merger signal is shown in Fig.\ref{fig:merger}-(a) and the time-series of signal in noise is shown in Fig.\ref{fig:merger}-(b). Figures \ref{fig:merger}-(c),-(d),-(e) show the sifted IMFs; in the 3rd IMF the signal can be seen, largely separate from the noise.

Figures \ref{fig:merger}-(f) and \ref{fig:merger}-(g) show the instantaneous frequency and power derived from the Hilbert Transform of IMF3. The signal frequency and power are clearly visible in the Hilbert Transform of IMF3: the frequency shows a ramp during the merger and levels off to a constant during the ringdown, while the power shows the expected rise and decay. This level of detail will aid signal identification, allowing comparison of signals from multiple detectors. This may be contrasted with the Fourier Transform of the merger, which would give the power of only 1 point with a 200~Hz frequency spread for the entire 5~msec. Finally, we note that further noise reduction through averaging of the IF and IA over time is possible, as they are oversampled: the plots of Fig.\ref{fig:merger} are sampled at the LIGO data rate of 16~kHz while the frequencies of interest for LIGO analysis are typically below 1~kHz.

In analogy with an FFT-based search algorithm that looks for excess power in the Fourier power spectrum \cite{Anderson2001}, the HHT can be used to search for excess power in the time domain. For a given time record, the summed power is computed, and compared to the background level; thus clear evidence of a signal is seen in Fig.\ref{fig:merger}-(g), the power associated with IMF3. In comparison to an FFT-based search, the HHT is most effective for short signals ($< 20{\rm msec}$), where FFT analysis tends to lose sensitivity from time-frequency spreading. Thus the HHT is a useful tool, for example, in searching for black hole binary merger signals up to 100 solar masses.

{\it Analysis of Spectral Shapes ---}
To facilitate the analysis of data spectral shapes, we use the HHT analogy of the power spectrum, called the "marginal spectrum." The marginal spectrum is defined as:
\begin{equation}
M(\omega)=\sum_i\int_0^{T}|a_i(\omega,t)|^2 dt
\end{equation}
where $a_i(\omega,t)$ is the amplitude as a function of the instantaneous frequency at a given point in time and $i$ is an IMF summation index. Thus the marginal spectrum is a measure of the total power present at a frequency $\omega$ over the time interval $T$.

\begin{figure}
\includegraphics*[width=\linewidth]{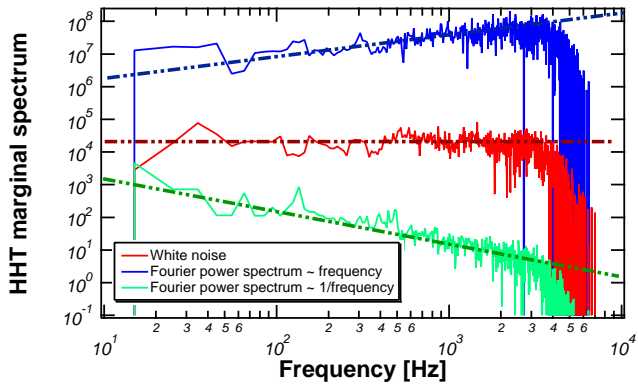}
\caption{\label{fig:spectrum}HHT marginal spectrum of $1/16{\rm sec}$ time-series with Fourier spectral shapes that are flat, proportional to frequency, and proportional to 1/frequency.}
\end{figure}

Figure \ref{fig:spectrum} shows the marginal spectrum of a time-series of length 1/16 second with the following Fourier power spectral shapes: flat, proportional to frequency, and inversely proportional to frequency. The marginal spectra are seen to have the expected shapes. Since the shape of the marginal spectrum does not depend on the integration time, it may be used to examine the detector noise stationarity in fine detail.

{\it Application of HHT to GW detector characterization ---}
The LIGO detector includes a large number of mechanical resonators, servos, cavity optical resonances, and light and sideband frequencies with differing Q's and transient excitation levels \cite{Abbott2004}. They can produce signals which can interact with each other in very transient and non-linear ways, for example upconverting noise at low frequencies into the signal band \cite{Camp2000}. The HHT can be used to examine the detailed dynamics of their interactions in a way that is unavailable to Fourier decomposition.

\begin{figure}
\includegraphics*[width=\linewidth]{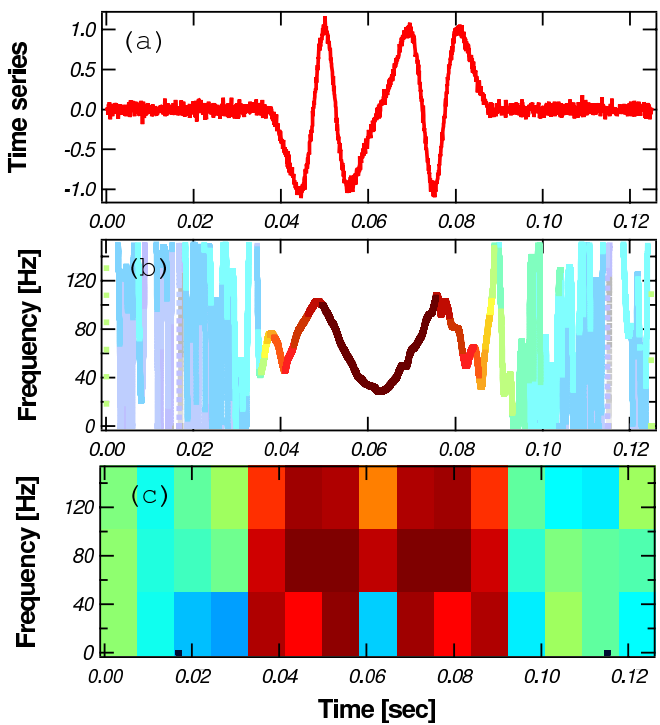}
\caption{\label{fig:TFplot}Time-frequency analysis of simulated instrumental artifact (Eq.(\ref{eqn:simsignal})). a) Time-series of frequency modulated, 40~msec transient in white noise. b) IF and IA from HHT: the modulation of the 60~Hz carrier from 20~Hz to 100~Hz is clearly visible. c) FFT-based spectrogram: time-frequency spreading blurs the modulation so that detail of its time structure is lost.}
\end{figure}

We show the capability of the HHT to resolve the following non-linear, transient waveform:
\begin{equation}
y=\cos{[2\pi 60 t + \sin{(2\pi 40 t)}]}
\label{eqn:simsignal}
\end{equation}
which contains a 60~Hz carrier frequency, phase modulated at 40~Hz. This could be produced by a laser frequency variation at 60~Hz coupled to seismic isolation stack motion at 40~Hz, for example. This waveform, which lasts only 40~msec, is shown added to white noise in Fig.\ref{fig:TFplot}-(a).

After application of the HHT, the modulation of the carrier frequency from 20~Hz to 100~Hz is clearly visible in Fig.\ref{fig:TFplot}-(b), where the IA is shown as a function of color. In Fig.\ref{fig:TFplot}-(b) the process of the upconversion of noise components at 40 and 60~Hz (outside the LIGO signal band), into components at 100~Hz (within the signal band), is clearly revealed. Resolution at this level will aid in understanding transient experimental artifacts and distinguishing them from real signals. In contrast, Fig.\ref{fig:TFplot}-(c) shows the FFT-based spectrogram, where time-frequency spreading blurs the modulation. In this figure the detail of the time structure of the modulation is lost.

{\it Summary ---}
We have presented the application of the Hilbert-Huang Transform to the analysis of GW data. The high time-frequency resolution of the HHT, which is not limited by time-frequency spreading, will have important application to both signal detection and detector characterization.

We acknowledge helpful comments on this manuscript by Malik Rakhmanov, Peter Shawhan, and Steve Ritz, and early programming assistance from Sean McWilliams.

\end{document}